%% file: main.tex
\documentclass[10pt,twocolumn,letterpaper]{article}

\usepackage{cvpr}
\usepackage{times}
\usepackage{epsfig}
\usepackage{graphicx}
\usepackage{amsmath}
\usepackage{amssymb}

\usepackage{hhline}
\usepackage{array}
\usepackage{booktabs}
\usepackage{multirow}
\usepackage{fancyhdr}

\usepackage[breaklinks=true,bookmarks=false]{hyperref}

\cvprfinalcopy 

\def\cvprPaperID{****} 

\newcommand{\mbf}{\mathbf}
\begin{document}

\title{Learning to dance: A graph convolutional adversarial network to generate realistic dance motions from audio}

\author{Jo\~ao P. Ferreira\\
UFMG\\
\and
Thiago M. Coutinho\\
UFMG\\
\and
Thiago L. Gomes\\
UFOP\\
\and
Jos\'e F. Neto\\
UFMG\\
\and
Rafael Azevedo\\
UFMG\\
\and
Renato Martins\\
INRIA\\
\and
Erickson R. Nascimento\\
UFMG\\
}

\maketitle
\thispagestyle{fancy}

\input{abstract}

\input{introduction}

\input{related_work.tex}

\input{methodology.tex}

\input{dataset.tex}
\input{experiments-and-results.tex}

\input{conclussion.tex}

{\small
\bibliographystyle{ieee_fullname}
\bibliography{bibliografia}
}

\end{document}

%% file: abstract.tex
\begin{abstract}

    Synthesizing human motion through learning techniques is becoming an increasingly popular approach to alleviating the requirement of new data capture to produce animations. Learning to move naturally from music, \ie, to dance, is one of the more complex motions humans often perform effortlessly. Each dance movement is unique, yet such movements maintain the core characteristics of the dance style. Most approaches addressing this problem with classical convolutional and recursive neural models undergo training and variability issues due to the non-Euclidean geometry of the motion manifold structure. 
In this paper, we design a novel method based on graph convolutional networks to tackle the problem of automatic dance generation from audio information. Our method uses an adversarial learning scheme conditioned on the input music audios to create natural motions preserving the key movements of different music styles. We evaluate our method with three quantitative metrics of generative methods and a user study. The results suggest that the proposed GCN model outperforms the state-of-the-art dance generation method conditioned on music in different experiments. Moreover, our graph-convolutional approach is simpler, easier to be trained, and capable of generating more realistic motion styles regarding qualitative and different quantitative metrics. It also presented a visual movement perceptual quality comparable to real motion data. The dataset and project are publicly available at: \url{https://www.verlab.dcc.ufmg.br/motion-analysis/cag2020}.

    \end{abstract}
    
        %

%% file: introduction.tex
\section{Introduction}



One of the enduring grand challenges in Computer Graphics is to provide plausible animations to virtual avatars. Humans have a large set of different movements when performing activities such as walking, running, jumping, or dancing. Over the past several decades, modeling such movements has been delegated to motion capture systems. Despite remarkable results achieved by highly skilled artists using captured motion data, the human motion has a rich spatiotemporal distribution with an endless variety of different motions. Moreover, human motion is affected by complex situation-aware aspects, including the auditory perception, physical conditions such as the person's age and its gender, and cultural background.

Synthesizing motions through learning techniques is becoming an increasingly popular approach to alleviating the requirement of capturing new real motion data to produce animations. The motion synthesis has been applied to a myriad of applications such as graphic animation for entertainment, robotics, and multimodal graphic rendering engines with human crowds~\cite{ikeuchi2018describing}, to name a few. Movements of each human being can be considered unique having its particularities, yet such movements preserve the characteristics of the motion style (\eg, walking, jumping, or dancing), and we are often capable of identifying the style effortlessly. When animating a virtual avatar, the ultimate goal is not only retargeting a movement from a real human to a virtual character but embodying motions that resemble the original human motion. In other words, a crucial step to achieve plausible animation is to learn the motion distribution and then draw samples (\ie, new motions) from it. For instance, a challenging human movement is dancing, where the animator does not aim to create avatars that mimic real poses but to produce a set of poses that match the music's choreography, while preserving the quality of being individual.

\begin{figure}[t!]
    \centering
    \includegraphics[width=0.9\linewidth]{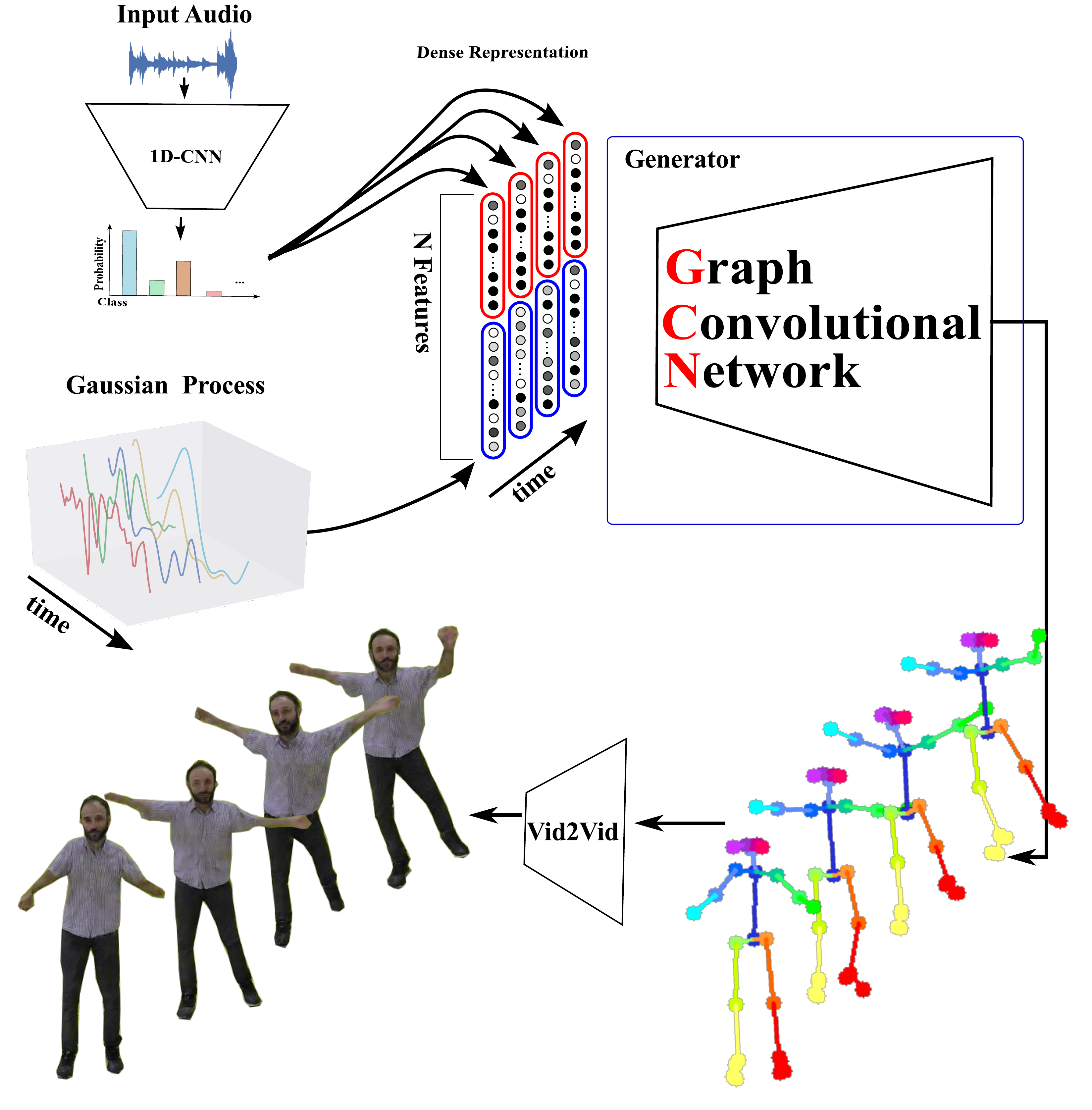}
    \caption{Our approach is composed of three main steps: First, given a music sound as input, we classify the sound according to the dance style; Second, we generate a temporal coherent latent vector to condition the motion generation, \ie, the spatial and temporal position of joints that define the motion; Third, a generative model based on a graph convolutional neural network is trained in an adversarial manner to generate the sequences of human poses. To exemplify an application scenario, we render  animations of virtual characters performing the motion generated by our method.}
    \label{fig:teaser}
\end{figure}

In this paper, we address the problem of synthesizing dance movements from music using adversarial training and a convolutional graph network architecture (GCN). Dancing is a representative and challenging human motion. Dancing is more than just performing pre-defined and organized locomotor movements, but it comprises steps and sequences of self-expression. In dance moves, both the particularities of the dancer and the characteristics of the movement play an essential role in recognizing the dance style. Thus, a central challenge in our work is to synthesize a set of poses taking into account three main aspects: firstly, the motion must be plausible, \ie, a blind evaluation should present similar results when compared to real motions; secondly, the synthesized motion must retain all the characteristics present in a typical performance of the music's choreography; third, each new set of poses should not be strictly equal to another set, in other words, when generating a movement for a new avatar, we must retain the quality of being individual. Figure~\ref{fig:teaser} illustrates our methodology.

Creating motions from sound relates to the paradigm of embodied music cognition. It couples perception and action, physical environmental conditions, and subjective user experiences (cultural heritage)~\cite{leman2015role}. Therefore, synthesizing realistic human motions regarding embodying motion aspects remains a challenging and active research field~\cite{ginosar2019gestures,yan2019convolutional}. Modeling distributions over movements is a powerful tool that can provide a large variety of motions while not removing the individual characteristics of each sample that is drawn. Furthermore, by conditioning these distributions, for instance, using an audio signal like music, we can select a sub-population of movements that match with the input signal. Generative models have demonstrated impressive results in learning data distributions. These models have been improved over the decades through machine learning advances that broadened the understanding of learning models from data. In particular, advances in the deep learning techniques yielded an unprecedented combination of effective and abundant techniques able to predict and generate data. The result was an explosion in highly accurate results in tasks of different fields. The explosion was felt first and foremost in the Computer Vision community. From high accuracy scores in image classification using convolutional neural networks (CNN) to photo-realistic image generation with the generative adversarial networks (GAN)~\cite{goodfellow2014generative}, Computer Vision field has been benefited with several improvements in the deep learning methods. Both Computer Vision and Computer Graphics fields also achieved significant advances in processing multimodal data present in the scene by using several types of sensors. These advances are assigned to the recent rise of learning approaches, especially convolutional neural networks. Also, these approaches have been explored to synthesize data from multimodal sources, and the audio data is one that is achieving the most impressive results, as the work presented by \cite{cudeiro2019capture}.

Most recently, networks operating on graphs have emerged as promising and effective approaches to deal with problems which structure is known {\it a priori}. A representative approach is the work of Kipf and Welling~\cite{kipf2016semi}, where a convolutional architecture that operates directly on graph-structured data is used in a semi-supervised classification task. Since graphs are natural representations for the human skeleton, several approaches using GCN have been proposed in the literature to estimate and generate human motion. Yan~\etal~\cite{yan2019convolutional}, for instance, presented a framework based on GCNs that generates a set of skeleton poses by sampling random vectors from a Gaussian process (GP). Despite being able to create sets of poses that mimic a person's movements, the framework does not provide any control over the motion generation. As stated, our methodology synthesizes human movements also using GCN, but unlike Yan~\etal's work, we can control the style of the movement using audio data while preserving the plausibility of the final motions. 
We argue that movements of a human skeleton, which has a graph-structured model, follow complex sequences of poses that are temporal related, and the set of defined and organized movements can be better modeled using a convolutional graph network trained using adversarial regime. 

In this context, we propose an architecture that manages audio data to synthesize motion. Our method starts encoding a sound signal to extract the music style using a CNN architecture. The music style and a spatial-temporal latent vector are used to condition a GCN architecture that is trained in an adversarial regime to predict 2D human body joint positions over time. Experiments with a user study and quantitative metrics showed that our approach outperforms the state-of-the-art method and provides plausible movements while maintaining the characteristics of different dance styles.

The contribution of this paper can be summarized as follows:
\begin{itemize}
    \item A new conditional GCN architecture to synthesize human motion based on auditory data. In our method, we push further the adversarial learning to provide multimodal data learning with temporal dependence;
    \item A novel multimodal dataset with paired audio, motion data and videos of people dancing different music styles. 
\end{itemize}

%% file: related_work.tex
\section{Related Work}

\paragraph{\textbf{Sound and Motion}} 
Recently, we have witnessed an overwhelming growth of new approaches to deal with the tasks of transferring motion style and building animations of people from sounds. 
For example, Bregler~\etal~\cite{BreglerCS97} create videos of a subject saying a phrase they did not speak originally, by reordering the mouth images in the training input video to match the phoneme sequence of the new audio track. 
In the same direction, Weiss~\cite{Weiss05} applied a data-driven multimodal approach to produce a 2D video-realistic audio-visual ``Talking Head'', using F0 and Mel-Cepstrum coefficients as acoustical features to model the speech.
Aiming to synthesize human motion according to music characteristics such as rhythm, speed, and intensity, Shiratori and Ikeuchi~\cite{shiratori2008synthesis} established keyposes according to changes in the rhythm and performer's hands, feet, and center of mass. Then, they used music and motion feature vectors to select candidate motion segments that match the music and motion intensity. Despite the impressive results, the method fails when the keyposes are in fast segments of the music. 

Cudeiro~\etal~\cite{cudeiro2019capture} presented an encoder-decoder network that uses audio features extracted from DeepSpeech~\cite{hannun2014deep}. The network generates realistic 3D facial animations conditioned on subject labels to learn different individual speaking styles. To deform the human face mesh, Cudeiro~\etal~encode the audio features in a low-dimensional embedding space. Although their model is capable of generalizing facial mesh results for unseen subjects, they reported that the final animations were distant from the natural captured real sequences. Moreover, the introduction of a new style is cumbersome since it requires a collection of 4D scans paired with audios.  Ginosar~\etal~\cite{ginosar2019gestures} enable translation from speech to gesture, generating arms and hand movements by mapping audio to pose. They used an adversarial training, where a U-Net architecture transforms the encoded audio input into a temporal sequence of 2D poses. In order to produce more realistic results, the discriminator is conditioned on the differences between each pair of subsequently generated poses. However, their method is subject-specific and does not generalize to other speakers.


More related work to ours is the approach proposed by Lee~\etal~\cite{lee2019dancing2music}. The authors use a complex architecture to synthesize dance movements (expressed as a sequence of 2D poses) given a piece of input music.
Their architecture is based on an elaborated decomposition-to-composition framework trained with an adversarial learning scheme. 
Our graph-convolutional based approach, on its turn, is simpler, easier to be trained, and generates more realistic motion styles regarding qualitative and different quantitative metrics. 

\paragraph{\textbf{Generative Graph Convolutional Networks}} 
 Since the seminal work of Goodfellow~\etal~\cite{goodfellow2014generative}, generative adversarial networks (GAN) have been successfully applied to a myriad of hard problems, notably for the synthesis of new information, such as of images~\cite{Karras2018ICLR}, motion~\cite{chan2019dance}, and pose estimation~\cite{posenet2017}, to name a few. Mirza and Osindero~\cite{mirza2014conditional} proposed Conditional GANs (cGAN), which provides some guidance into the data generation. Reed~\etal~\cite{pmlr-v48-reed16} synthesize realistic images from text, demonstrating that cGANs can also be used to tackle multi-modal problems. 
Graph Convolutional Networks (GCN) recently emerged as a powerful tool for learning from data by leveraging geometric properties that are embedded beyond n-dimensional Euclidean vector spaces, such as graphs and simplicial complex. In our context, conversely to classical CNNs, GCNs can model the motion manifold space structure~\cite{Jain_2016_CVPR, yan2018spatial, yan2019convolutional}. 
Yan~\etal~\cite{yan2018spatial} applied GCNs to model human movements and classify actions. After extracting 2D human poses for each frame from the input video, the skeletons are processed by a Spatial-Temporal Graph Convolutional Network (ST-GCN). Yan~\etal~proceeded in exploiting the representation power of GCNs and presented the Convolutional Sequence Generation Network (CSGN)~\cite{yan2019convolutional}. By sampling correlated latent vectors from a Gaussian process and using temporal convolutions, the CSGN architecture was capable of generating temporal coherent long human body action sequences as skeleton graphs. 
Our method takes one step further than~\cite{yan2018spatial,yan2019convolutional}. It generates human skeletal-based graph motion sequences conditioned on acoustic data, \ie, music. By conditioning the movement distributions, our method learns not only creating plausible human motion, but it also learns the music style signature movements from different domains.


\paragraph{\textbf{Estimating and Forecasting Human Pose}}

Motion synthesis and motion analysis problems have been benefited from the improvements in the accuracy of human pose estimation methods. Human pose estimation from images, for its turn, greatly benefited from the recent emergence of large datasets~\cite{COCO_dataset, MPII_dataset,guler2018densepose} with annotated positions of joints, and dense correspondences from 2D images to 3D human shapes~\cite{openpose,ap_li2018crowdpose,ap_xiu2018poseflow,guler2018densepose,kolotouros2019spin,humanMotionKanazawa19,kocabas2019vibe}.  
This large amount of annotated data has made possible important milestones towards predicting and modeling human motions~\cite{wang20143d,Gui_2018_ECCV,ghosh2017learning,fragkiadaki2015recurrent,wang2019spatio}. 
The recent trend in time-series prediction with recurrent neural networks (RNN) became popular in several frameworks for human motion prediction~\cite{fragkiadaki2015recurrent, martinez2017human,ghosh2017learning}. Nevertheless, the pose error accumulation in the predictions allows mostly predicting over a limited range of future frames~\cite{Gui_2018_ECCV}. Gui~\etal~\cite{Gui_2018_ECCV} proposed to overcome this issue by applying adversarial training using two global recurrent discriminators that simultaneously validate the sequence-level plausibility of the prediction and its coherence with the input sequence. Wang~\etal~\cite{wang2019spatio} proposed a network architecture to model the spatial and temporal variability of motions through a spatial component for feature extraction. Yet, these RNN models are known to be difficult to train and computationally cumbersome~\cite{Pascanu_2013}. As also noted by~\cite{lee2019dancing2music}, motions generated by RNNs tend to collapse to certain poses regardless of the inputs.

\paragraph{\textbf{Transferring Style and Human Motion}}
Synthesizing motion with specific movement style has been studied in a large body of prior works~\cite{smith2019efficient,2018-TOG-SFV,wang2018vid2vid,chan2019dance,Gomes_2020_WACV}. Most methods formulate the problem as transferring a specific motion style to an input motion~\cite{xia2015realtime,smith2019efficient}, or transferring the motion from one character to another, commonly referred as motion retargeting~\cite{Gleicher,motion_retargetting_1,Villegas_2018_CVPR}. Recent approaches explored deep reinforcement learning to model physics-based locomotion with a specific style~\cite{2017-TOG-deepLoco,Liu_Hodgins_2018,2018-TOG-SFV}. Another active research direction is transferring motion from video-to-video~\cite{wang2018vid2vid,chan2019dance,Gomes_2020_WACV}. However, the generation of stylistic motion from audio is less explored, and it is still a challenging research field.
Villegas~\etal~\cite{villegas2017learning} presented a video generation method based on high-level structure extraction, conditioning the creation of new frames on how this structure evolves in time, therefore preventing pixel-wise error prediction accumulation. Their approach was employed on long-term video prediction of humans performing actions by using 2D human poses as high-level structures. 

Wang~\etal~\cite{wang2018adversarial} discussed how adversarial learning could be used to generate human motion by using a sequence of autoencoders. The authors focused on three tasks: motion synthesis, conditional motion synthesis, and motion style transfer. As our work, their framework enables conditional movement generation according to a style label parameterization, but there is no multimodality associated with it. Jang~\etal~\cite{motion-sequential-20} presented a method inspired by sequence-to-sequence models to generate a motion manifold.
As a significant drawback, the performance of their method decreases when creating movements longer than 10s, which makes the method inappropriate to generate long sequences. Our approach, on the other hand, can create long movement sequences conditioned on different music styles, by taking advantage of the adversarial GCN's power to generate new long, yet recognizable, motion sequences.

%% file: methodology.tex
\section{Methodology}

\begin{figure}[t!]
    \centering
    \includegraphics[width=0.85\linewidth]{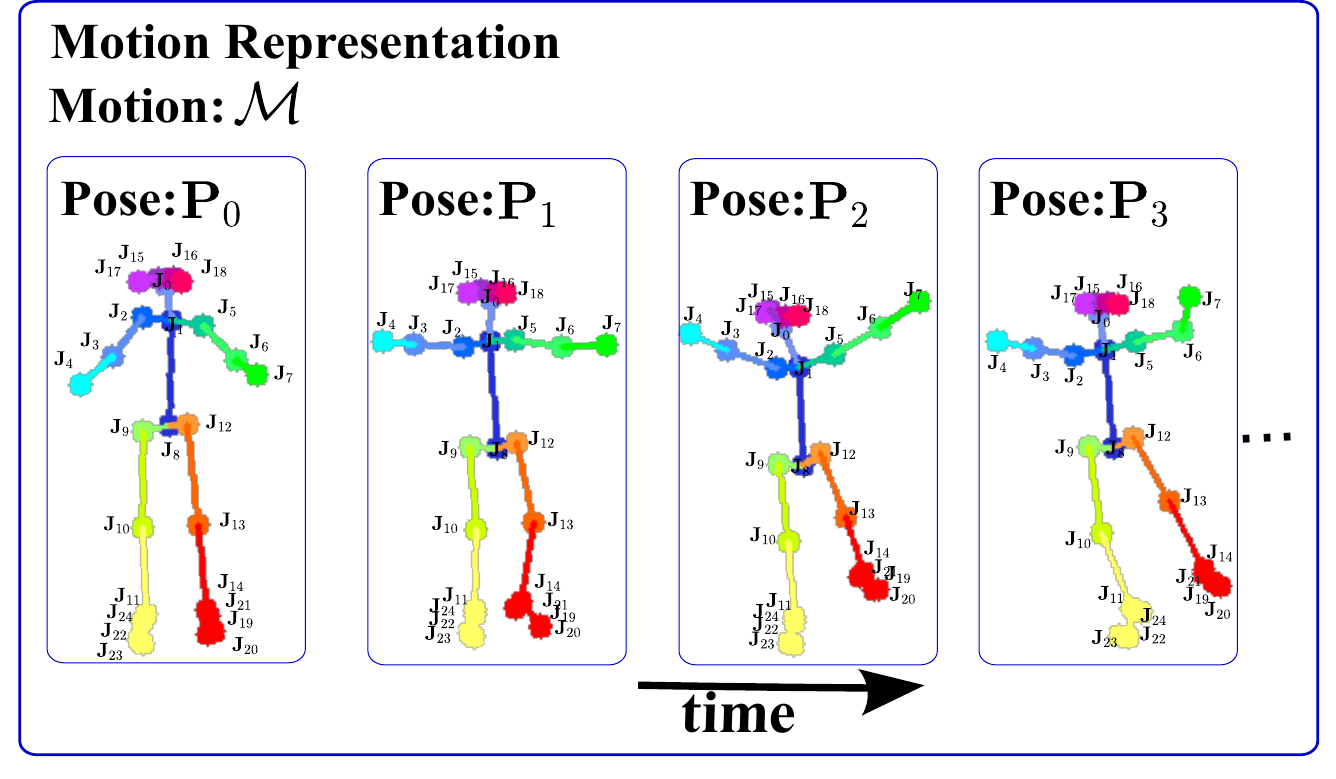}
   \caption{Motion and skeleton notations. In our method, we used a skeleton with $25$ 2D joints.}
    \label{fig:notations}
\end{figure}

Our method has been designed to synthesize a sequence of 2D human poses resembling a human dancing according to a music style. Specifically, we aim to estimate a motion $\mathcal{M}$ that provides the best fit for a given input music audio. $\mathcal{M}$ is a sequence of $N$ human body poses defined as:
\begin{align}
    \label{eq:motion}
        \mathcal{M} = \left [ \mbf P_0, \mbf P_1, \cdots, \mbf P_N \right ] \in \mathbb{R}^{N\times 25 \times 2},
\end{align}
\noindent where ${\mbf P_t = \left [ \mbf J_0, \mbf J_1, \cdots, \mbf J_{24} \right ]}$ is a graph representing the body pose in the frame $t$ and $\mbf J_i \in \mathbb{R}^2$ the 2D image coordinates of $i$-th node of this graph (see Figure~\ref{fig:notations}).

\begin{figure*}[!t]
    \centering
     \includegraphics[width=0.85\textwidth]{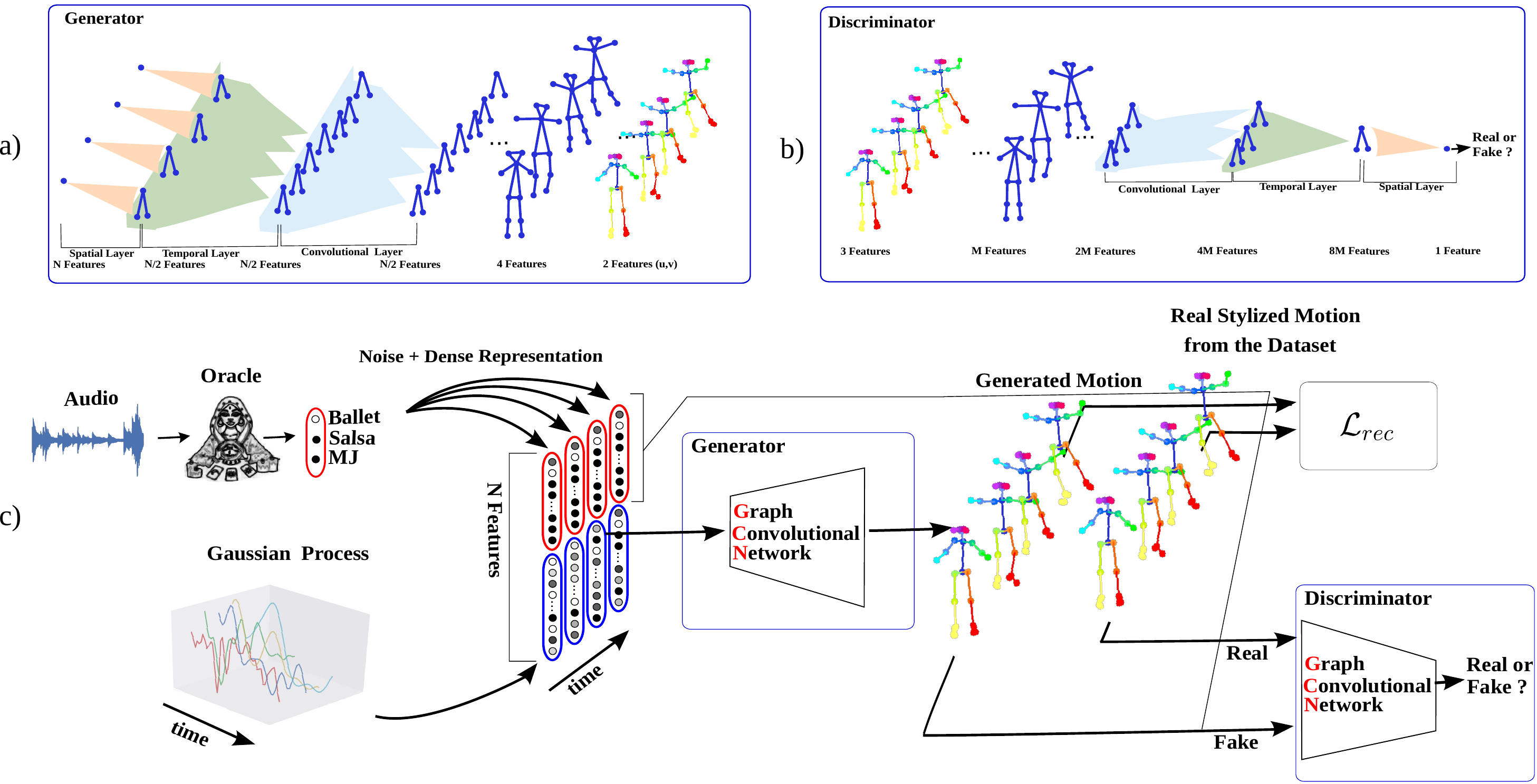}
    \caption{Motion synthesis conditioned on the music style. (a) GCN Motion Generator $G$; (b) GCN Motion Discriminator $D$; and (c) an overview of the adversarial training regime.}
    \label{fig:method}
\end{figure*}

Our approach consists of three main components, outlined in Figure~\ref{fig:method}. We start training a 1D-CNN classifier to define the input music style.
Then, the result of the classification is combined with a spatial-temporal correlated latent vector generated by a Gaussian process (GP). The GP allows us to sample points of Gaussian noise from a distribution over functions with a correlation between points sampled for each function. Thus, we can draw points from functions with different frequencies. This variation in the signal frequency enables our model to infer which skeleton joint is responsible for more prolonged movements and explore a large variety of poses. The latent vector aims at maintaining spatial coherence of the motion for each joint overtime. At last, we perform the human motion generation from the latent vector. In the training phase of the generator, we use the latent vector to feed a graph convolutional network that is trained in an adversarial regime on the dance style defined by an oracle algorithm. In the test phase, we replace the oracle by the 1D-CNN classifier. Thus, our approach has two training stages: \emph{i)} The training of the audio classifier to be used in the test phase and \emph{ii)} The GCN training with an adversarial regime that uses the music style to condition the motion generation.

\subsection{Sound Processing and Style Feature Extraction}

Our motion generation is conditioned by a latent vector that encodes information from the music style. 
In this context, we used the SoundNet~\cite{aytar2016soundnet} architecture as the backbone to a one-dimensional CNN. The 1D-CNN receives a sound in waveform and outputs the most likely music style considering three classes. The classifier is trained in a dataset composed of $107$ music files and divided into three music-dance styles: {\it Ballet}, {\it Salsa}, and {\it Michael Jackson (MJ)}. 

To find the best hyperparameters, we ran a $10$-fold cross-validation and kept the best model to predict the music style to condition the generator. Different from works~\cite{arandjelovic2018objects,vggish} that require 2D pre-processed sound spectrograms, our architecture is one-dimensional and works directly in the waveform. 
 




\subsection{Latent Space Encoding for Motion Generation}\label{sec:latent-code}


In order to create movements that follow the music style, while keeping particularities of the motion and being temporally coherent, we build a latent vector that combines the extracted music style with a spatiotemporal correlated signal from a Gaussian process.  It is noteworthy that our latent vector differs from the work of Yan~\etal~\cite{yan2019convolutional}, since we condition our latent space using the information provided by the audio classification. The information used to condition the motion generation, and to create our latent space, is a trainable dense feature vector representation of each music style. The dense music style vector representation works as a categorical dictionary, which maps a dance style class to a higher dimensional space.



Then, we combine a temporal coherent random noise with the music style representation in order to generate coherent motions over time. Thus, the final latent vector is the result of concatenating the dense trainable representation of the audio class with the coherent temporal signal in the dimension of the features. This concatenation plays a key role in the capability of our method to generate synthetic motions with more than one dancing style when the audio is a mix of different music styles. In other words, unlike a vanilla conditional generative model, which conditioning is limited to one class, we can condition over several classes over time. 

The coherent temporal signals are sampled from Radial Basis Function kernel (RBF)~\cite{rasmussen2003gaussian} to enforce temporal relationship among the $N$ frames. A zero-mean Gaussian process with a covariance function $\kappa$ is given by $(z_{t}^{(c)})\sim GP(0,\kappa)$, where $(z_{t}^{(c)})$ is the $c$-th component of $z_t$. The signal comprises $c$ functions with $t \in \mathbb{R}^{N/16}$ temporally coherent values. This provides a signal with a shape of ($C,T,V$), where $C$ is interpreted as the channels (features) of our graph, $T$ is related to the length of the sequence we want to generate, and $V$ is the spatial dimension of our graph signal. The covariance function $\kappa$ is defined as:
\begin{equation}
    \kappa(t,t') = \exp \left(-\frac{|t-t'|^2}{2\sigma_{c}^{2}}\right).
\end{equation}
\noindent In our tests, we used $C=512,T=4,V=1$ and ${\sigma_{c} = \sigma \left(\frac{c_{i}}{C}\right)}$, where $\sigma = 200$ was chosen empirically and $c_{i}$ varies for every value from $0$ to $C$.


The final tensor representing the latent vector has the size $(2C,T,V)$, where the sizes of $C$ and $T$ are the same as the coherent temporal signal. Note that the length of the final sequence is proportional to $T$ used in the creation of the latent vector. The final motion, after propagation in the motion generator, will have $16T = N$ frames; thus, we can generate samples for any FPS and length by changing the dimensions of the latent vector. Moreover, as the dimensions of the channels condition the learning, we can change the conditioning dance style over time.

The Gaussian process generates our random noise $z$ and the dense representation of the dance style is the variable used to condition our model $y$. The combination of both data is used as input for the generator.

\subsection{Conditional Adversarial GCN for Motion Synthesis}

To generate realistic movements, we use a graph convolutional neural network (GCN) trained with an adversarial strategy. The key idea in adversarial conditional training is to learn the data distribution while two networks compete against each other in a $minimax$ game. In our case, the motion generator $G$ seeks to create motion samples similar to those in the motion training set, while the motion discriminator $D$ tries to distinguish generated motion samples (fake) from real motions of the training dataset (real). Figure~\ref{fig:method} illustrates the training scheme. 

\paragraph{\textbf{Generator}}


The architecture of our generator $G$ is mainly composed of three types of layers: temporal and spatial upsampling operations, and graph convolutions. When using GCNs, one challenge that appears in an adversarial training is the requirement of upsampling the latent vector in the spatial and temporal dimensions to fit the motion space $\mathcal{M}$ (Equation~\ref{eq:motion}). 

\begin{figure}[t!]
    \centering
    \includegraphics[width=0.85\linewidth]{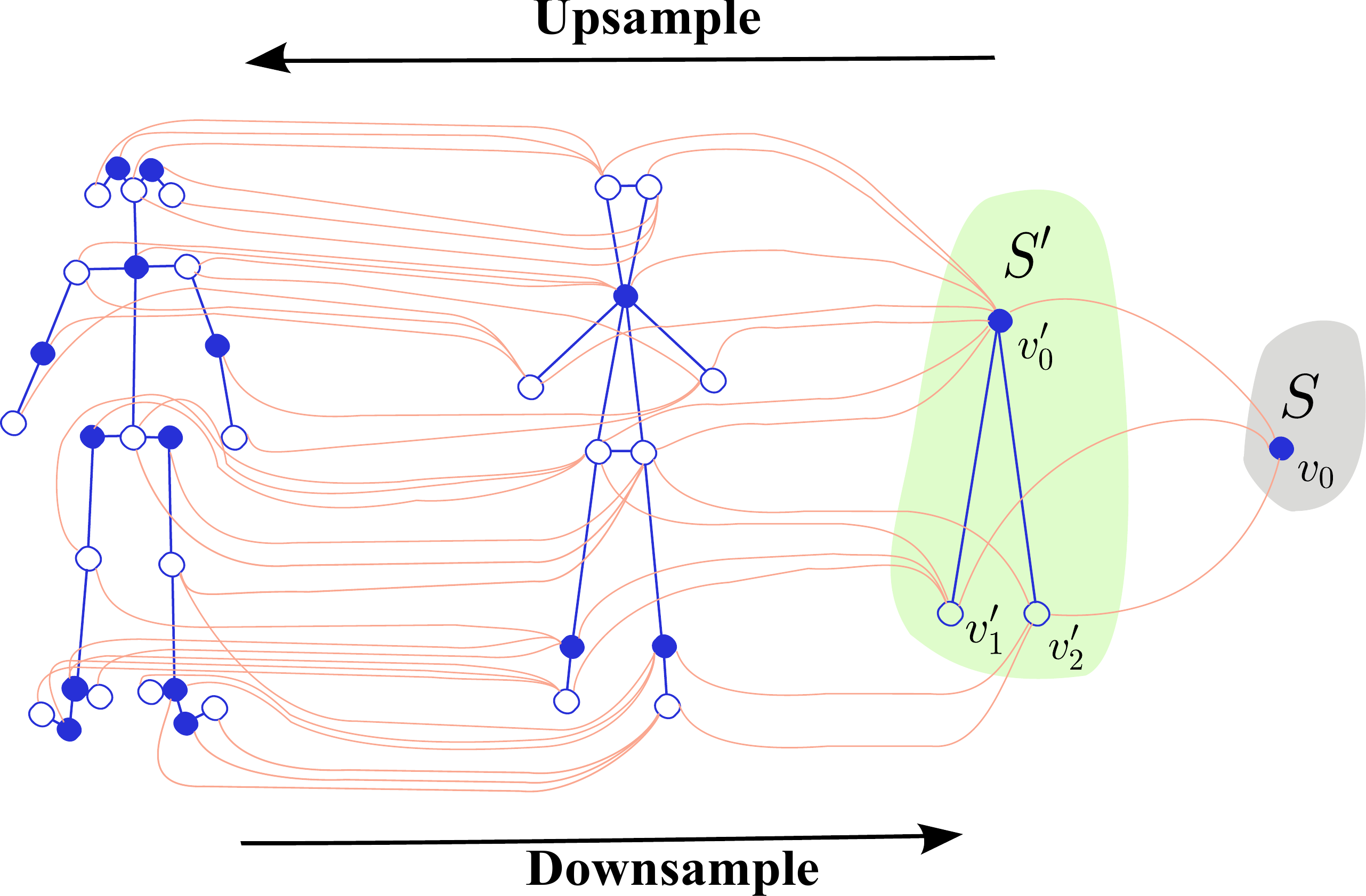}
   \caption{Graph scheme for upsampling and downsampling operations.}
    \label{fig:graph_op}
\end{figure}

The temporal upsampling layer consists of transposed 2D convolutions that double the time dimension, ignoring the input shape of each layer. Inspired by Yan~\etal~\cite{yan2019convolutional}, we also included in our architecture a spatial upsampling layer. This layer operates using an aggregation function defined by an adjacency matrix $A^\omega$ that maps a graph $S(V,E)$ with $V$ vertices and $E$ edges to a bigger graph $S'(V', E')$ (see Figure~\ref{fig:graph_op}). 
The network can learn the best values of $A^\omega$ that leads to a good upsampling of the graph by assigning different importance of each neighbor to the new set of vertices.

The first spatial upsampling layer starts from a graph with one vertex and then increases to a graph with three vertices. When creating the new vertices, the features $\mathbf{f_j}$ from the initial graph $S$ are aggregated by $A^\omega$ as follows:
\begin{equation}
    \mathbf{f'}_{i} = \sum_{k,j}A^\omega_{kij}\mathbf{f}_{j},
    \label{eq:upsampling}
\end{equation}
\noindent where $\mathbf{f'_i}$ contains the features of the vertices in the new graph $S'$ and $k$ indicates the geodesic distance between vertex $j$ and vertex $i$ in the graph $S'$. 

In the first layer of the generator, we have one node containing a total of $N$ features; these features represent our latent space (half from the Gaussian Process and a half from the audio representation). The features of the subsequent layers are computed by the operations of upsampling and aggregation. The last layer outputs a graph with $25$ nodes containing the $(x, y)$ coordinates of each skeleton joint.
For instance, in Figure~\ref{fig:graph_op} from right to left, we can see the upsampling operation, where we move from a graph with one vertex $S$ to a new graph containing three vertices $S'$. The aggregation function in $A^{\omega}$ is represented by the red links connecting the vertices between the graphs and the topology of graph $S'$. When $k=0$, vertex $v$ is directly mapped to vertex $v'$ (\ie, the distance between $v_{0}$ and $v'_{0}$ is $0$) and all values are zeros except the value of $i=0, j=0$ then $\mathbf{f'}_0= A^{\omega}_{0,0,0} \mathbf{f}_0$. Following the example, when $k=1$, we have $\mathbf{f'}_0 = A^{\omega}_{1,0,0} \mathbf{f}_0 $ e $\mathbf{f'}_1 = A^{\omega}_{1,1,0} \mathbf{f}_0$.

After applying the temporal and spatial upsampling operations, our generator uses the graph convolutional layers defined by Yan~\etal~\cite{yan2018spatial}. These layers are responsible for creating the spatio-temporal relationship between the graphs. Then, the final architecture comprises three sets of temporal, spatial, and convolutional layers: first, temporal upsampling for a graph with one vertex followed by an upsampling from one vertex to $3$ vertices, then one convolutional graph operation. We repeat these three operations for the upsampling from $3$ vertices to $11$, and finally from $11$ to $25$ vertices, which represents the final pose. Figure~\ref{fig:method}-(a) shows this GCN architecture.

\paragraph{\textbf{Discriminator}}

The discriminator $D$ has the same architecture used by the generator but using downsampling layers instead of upsampling layers. Thus, all transposed 2D convolutions are converted to standard 2D convolutions, and the spatial downsampling layers follow the same procedure of upsampling operations but using an aggregation matrix $B^\phi$ with trainable weights $\phi$, different from the weights learned by the generator. Since the aggregation is performed from a large graph $G'$ to a smaller one $G$, the final aggregation is given by
\begin{equation}
    \mathbf{f}_{i} = \sum_{k,j}B^\phi_{kij}\mathbf{f'}_{j}.
    \label{eq:downsampling}
\end{equation}

In the discriminator network, the feature vectors are assigned to each node as follows: the first layer contains a graph with $25$ nodes, where their feature vectors are composed of the $(x, y)$ coordinates on a normalized space and the class of the input motion. In the subsequent layers, the features of each node are computed by the operations of downsampling and aggregation. The last layer contains only one node that outputs the classification of the input data being fake or real. Figure~\ref{fig:method}-(b) illustrates the discriminator architecture.  




\paragraph{\textbf{Adversarial training}}

Given the motion generator and discriminator, our conditional adversarial network aims at minimizing the binary cross-entropy loss:
\begin{equation}
    \begin{split}
    \mathcal{L}_{cGAN}(G,D) = \min_{G}\max_{D}\left(\mathbb{E}_{x\sim p_{data}}(x)[\log D(x|y)]+ \right. \\ \left. \mathbb{E}_{z\sim p_{z}}(z)[\log (1 - D(G(z|y)))] \right),
    \end{split}
    \label{eq:cgan}
\end{equation}
\noindent where the generator aims to maximize the error of the discriminator, while the discriminator aims to minimize the classification fake-real error shown in Equation~\ref{eq:cgan}. In particular, in our problem, $p_{data}$ represents the distribution of real motion samples, $x = \mathcal{M}_{\tau}$ is a real sample from $p_{data}$, and $\tau \in [0-\mathcal{D}_{size}]$ and $\mathcal{D}_{size}$ is the number of real samples in the dataset. Figure~\ref{fig:method}-(c) shows a concise overview of the steps in our adversarial training.  

The latent vector, which is used by the generator to synthesize the fake samples $x'$, is represented by the variable $z$, the coherent temporal signal. The dense representation of the dance style is determined by $y$, and $p_{z}$, which is a distribution of all possible temporal coherent latent vectors generated by the Gaussian process. The data used by the generator $G$ in the training stage is the pair of temporal coherent latent vector $z$, with a real motion sample $x$, and the value of $y$ given by the music classifier that infers the dance style of the audio.

To improve the generated motion results, we use a motion reconstruction loss term applying $L_1$ distance in all skeletons over the $N$ motion frames as follows:
\begin{equation}
    \begin{split}
        \mathcal{L}_{rec} = \frac{1}{N}{\sum_{i=1}^{N} \mathcal{L}_{pose}\left(\mbf P_{t},\mbf P'_{t}\right)},
    \end{split}
\label{eq:loss-l1}
\end{equation}
\noindent with $\mbf P_t \in \mathcal{M}$ being the generated pose and $\mbf P'_t \in \mathcal{M'}$ a real pose from the training set and extracted with the OpenPose~\cite{openpose}. The pose distance is computed as $\mathcal{L}_{pose} = {\sum_{i=0}^{24}|\mbf J_{i} - \mbf J'_{i}|_1}/{25}$, following the notation shown in Equation~\ref{eq:motion}. 

Thus, our final loss is a weighted sum of the motion reconstruction and cGAN discriminator losses given by
\begin{equation}
     \mathcal L = \mathcal L_{cGAN} + \lambda \mathcal L_{rec},
    \label{eq:loss}
\end{equation}
\noindent where $\lambda$ weights the reconstruction term. The $\lambda$ value was chosen empirically, and was fixed throughout the training stage. The initial guess regarding the magnitude of $\lambda$ followed the values chosen by Wang~\etal~\cite{wang2018vid2vid}. 

We apply a cubic-spline interpolation in the final motion to remove eventual high frequency artifacts from the generated motion frames $\mathcal{M}$.

%% file: dataset.tex
\section{Audio-Visual Dance Dataset}
\begin{figure*}[!t]
    \centering
    \includegraphics[width=0.88\linewidth]{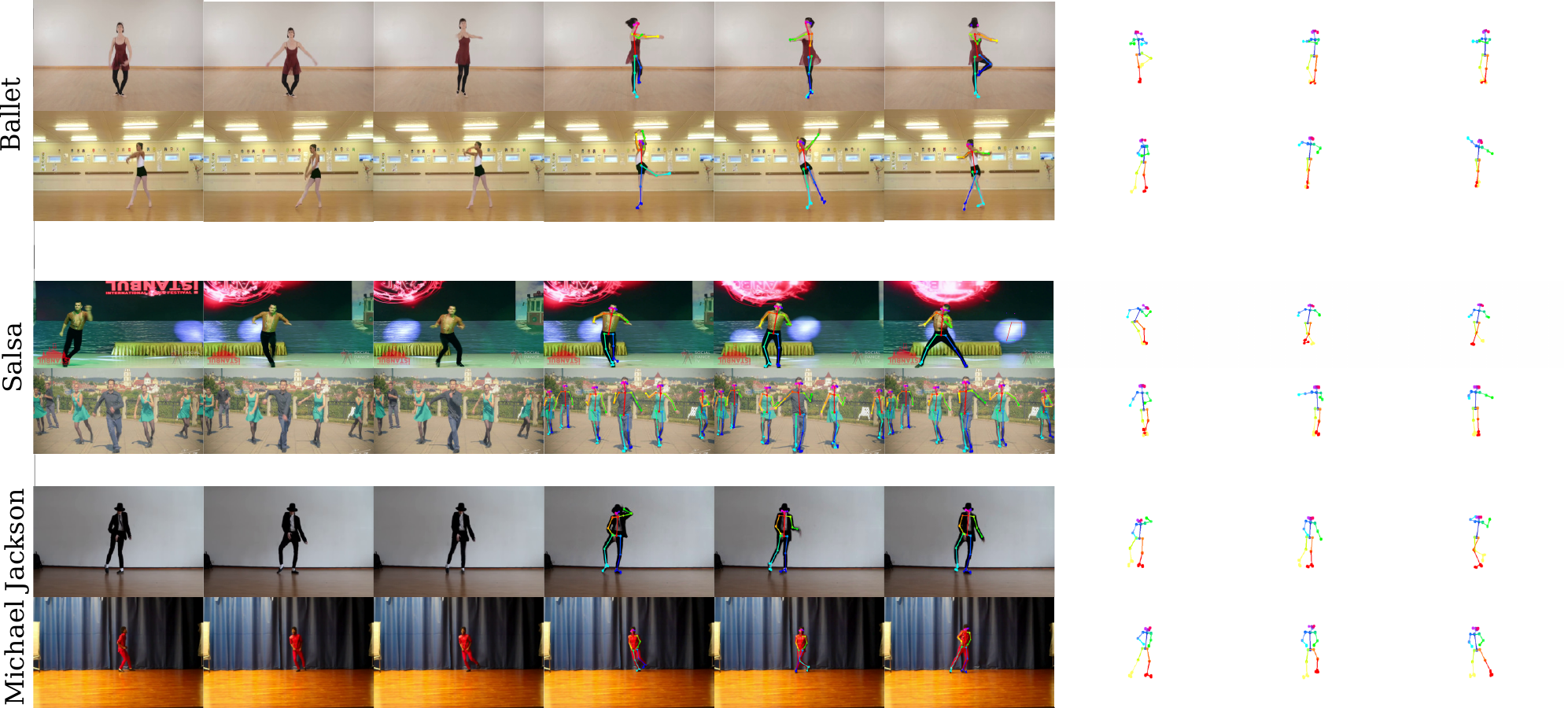}
    \caption{Video samples of the multimodal dataset with carefully annotated audio and 2D human motions of different dance styles.}

    \label{fig:dataset}
\end{figure*}
We build a new dataset composed of paired videos of people dancing different music styles. The dataset is used to train and evaluate the methodologies for motion generation from audio. We split the samples into training and evaluation sets that contain multimodal data for three music/dance styles: Ballet, Michael Jackson, and Salsa.  These two sets are composed of two data types: visual data from careful-selected parts of publicly available videos of dancers performing representative movements of the music style and audio data from the styles we are training. Figure~\ref{fig:dataset} shows some data samples of our dataset.


In order to collect meaningful audio information, several playlists from YouTube were chosen with the name of the style/singer as a search query. 
The audios were extracted from the resulting videos of the search and resampled to the standard audio frequency of 16KHz. For the visual data, we started by collecting videos that matched the music style and had representative moves. Each video was manually cropped in parts of interest, by selecting representative moves for each dance style present in our dataset. Then, we standardize the motion rate throughout the dataset and convert all videos to $24$ frames-per-second (FPS), maintaining a constant relationship between the number of frames and speed of movements of the actors. We annotate the $25$ 2D human joint poses for each video by estimating the pose with OpenPose~\cite{openpose}. Each motion sample is defined as a set of 2D human poses of $64$ consecutive frames.


To improve the quality of the estimated poses in the dataset, we handled the miss-detection of joints by exploiting the body dynamics in the video. Since abrupt motions are not expected in the joints in a short interval of frames, we recreate a missing joint and apply the transformation chain of its parent joint. In other words, we infer the missing-joint position of a child's joint by making it follow its parent movement over time. Thus, we can keep frames with a miss-detected joint on our dataset.


\subsection{Motion Augmentation}



We also performed motion data augmentation to increase the variability and number of motion samples. We used the Gaussian process described in Section~\ref{sec:latent-code} to add temporally coherent noise in the joints lying in legs and arms over time. Also, we performed temporal shifts (strides) to create new motion samples.
For the training set, we collected $69$ samples and applied the temporal coherent Gaussian noise and a temporal shift of size $32$.
In the evaluation set, we collected $229$ samples and applied only the temporal shift of size $32$ for Salsa and Ballet and $16$ for Michael Jackson because of the lower number of samples (see Table~\ref{tab:dataset}). The temporal Gaussian noise was not applied in the evaluation set. 
The statistics of our dataset are shown in Table~\ref{tab:dataset}. The resulting audio-visual dataset contains thousands of coherent video, audio, and motion samples that represent characteristic movements for the considered dance styles.\footnote{The dataset and project are publicly available at: \url{https://www.verlab.dcc.ufmg.br/motion-analysis/cag2020}.} 

We performed evaluations with the same architecture and hyperparameters, but without data augmentation, the performance on the Fréchet Inception Distance (FID) metric was worse than when using data augmentation. Moreover, we observed that the motions did not present variability, the dance styles were not well pictured, and in the worst cases, body movements were difficult to notice.

\begin{table}[t!]
    \centering
    \caption{Statistics of our dataset. The bold values are the number of samples used in the experiments.}
    \label{tab:dataset}
	\resizebox{\linewidth}{!}{
    \begin{tabular}{@{}crrrrrrrrrrrc@{}}
        \toprule
        & \multicolumn{4}{c}{Training Dataset} & \phantom{abc} & \multicolumn{4}{c}{Evaluation Dataset} \\
        \cmidrule{2-5} \cmidrule{7-10}
        \multicolumn{1}{l}{\textbf{Setup}} & \multicolumn{1}{c}{Ballet}       & \multicolumn{1}{c}{MJ} & \multicolumn{1}{c}{Salsa} &
        \multicolumn{1}{c}{Total} && \multicolumn{1}{c}{Ballet}       & \multicolumn{1}{c}{MJ} & \multicolumn{1}{c}{Salsa} &
        \multicolumn{1}{c}{Total}\\
        \midrule
        
        \multicolumn{1}{l}{w/o Data Augmentation} & $16$ & $26$ & $27$ & $69$ && $73$ & $30$ & $126$ & $229$\\
        \multicolumn{1}{l}{w/ Data Augmentation} & $525$ & $966$ & $861$ & $\mathbf{2,352}$ && $134$ & $102$ &$235$ & $\mathbf{471}$ \\
    
        \bottomrule
    \end{tabular}}
\end{table}

%% file: experiments-and-results.tex
\section{Experiments and Results}

To assess our method, we conduct several experiments evaluating different aspects of motion synthesis from audio information. We also compared our method to the state-of-the-art technique proposed by Lee~\etal~\cite{lee2019dancing2music}, hereinafter referred to as D2M. We choose to compare our method to D2M since other methods have major drawbacks that make a comparison with our method unsuitable, such as different skeleton structures in~\cite{ginosar2019gestures}.
Unfortunately, due to the lack of some components in the publicly available implementation of D2M, few adjustments were required in their audio preprocessing step. We standardized the input audio data by selecting the maximum length of the audio divisible by 28, defined as $L$, and reshaping it to a tensor of dimensions $\left(\frac{L}{28}, 28\right)$ to match the input dimensions of their architecture.

The experiments are as follows: \textit{i)} We performed a perceptual user study using a blind evaluation with users trying to identify the dance style of the dance moves. For a generated dance video, we ask the user to choose what style (Ballet, Michael Jackson (MJ), or Salsa) the avatar on the video is dancing; \textit{ii)} Aside from the user study, we also evaluated our approach on commonly adopted quantitative metrics in the evaluation of generative models, such as Fréchet Inception Distance (FID), GAN-train, and GAN-test~\cite{Shmelkov18}. 


\subsection{Implementation and Training Details}

\paragraph{\textbf{Audio and Poses Preprocessing}}

Our one-dimensional audio CNN was trained for $500$ epochs, with batch size equal to $8$, Adam optimizer with $\beta_1=0.5$ and $\beta_2=0.999$, and learning rate of $0.01$.
Similar to \cite{oord2016wavenet}, we preprocessed the input music audio using a $\mu-law$ non-linear transformation to reduce noise from audio inputs not appropriately recorded. We performed $10$-fold cross-validation to choose the best hyperparameters. 


In order to handle different shapes of the actors and to reduce the effect of translations in the 2D poses of the joints, we normalized the motion data used during the adversarial GCN training. We managed changes beyond body shape and translations, such as the situations of actors lying on the floor or bending forward, by selecting the diagonal distance of the bounding box encapsulating all 2D body joints $\mbf P_t$ of the frame as scaling factor. The normalized poses are given by: 
\begin{equation}
    \bar{\mbf J_i} = \frac{1}{\delta}\left(\mbf J_i - \left(\frac{\Delta u}{2}, \frac{\Delta v}{2}\right) \right) + 0.5,
    \label{eq:norm}
\end{equation} 
\noindent where $\delta = \sqrt{(\Delta u)^2 + (\Delta v)^2}$, and $(\Delta u, \Delta v)$ are the differences between right-top position and left-bottom position of the bounding box of the skeleton in the image coordinates $(u,v)$. 

\paragraph{\textbf{Training}}

We trained our GCN adversarial model for $500$ epochs. We observed that additional epochs only produced slight improvements in the resulting motions. 
In our experiments, we select $N=64$ frames, roughly corresponding to motions of three seconds at $24$ FPS. We select $64$ frames as the size of our samples to follow a similar setup presented in~\cite{ginosar2019gestures}. Moreover, the motion sample size in~\cite{lee2019dancing2music} also adopted motion samples of $32$ frames. In general, the motion sample size is a power of $2$, because of the nature of the conventional convolutional layers adopted in both~\cite{ginosar2019gestures,lee2019dancing2music}. 
However, it is worth noting that our method can synthesize long motion sequences. We use a batch size of $8$ motion sets of $N$ frames each. 
We optimized the cGAN with Adam optimizer for the generator with $\beta_1=0.5$ and $\beta_2=0.999$ with learning rate of $0.002$. The discriminator was optimized with Stochastic Gradient Descent (SGD) with a learning rate of $2\times10^{-4}$. We used $\lambda = 100$ in Equation~\ref{eq:loss}. Dropout layers were used on both generator and discriminator to prevent overfitting.

\paragraph{\textbf{Avatar Animations}}

As an application of our formulation, we animate three virtual avatars using the generated motions to different music styles. The image-to-image translation technique vid2vid~\cite{wang2018vid2vid} was selected to synthesize videos. We trained vid2vid to generate new images for these avatars, following the multi-resolution protocol described in~\cite{wang2018vid2vid}. 
For inference, we feed vid2vid with the output of our GCN. We highlight that any motion style transfer method can be used with few adaptations, as for instance, the works of~\cite{chan2019dance,Gomes_2020_WACV}. 




\subsection{User Study}


\begin{figure*}[t!]
    \centering
    \includegraphics[width=0.9\linewidth]{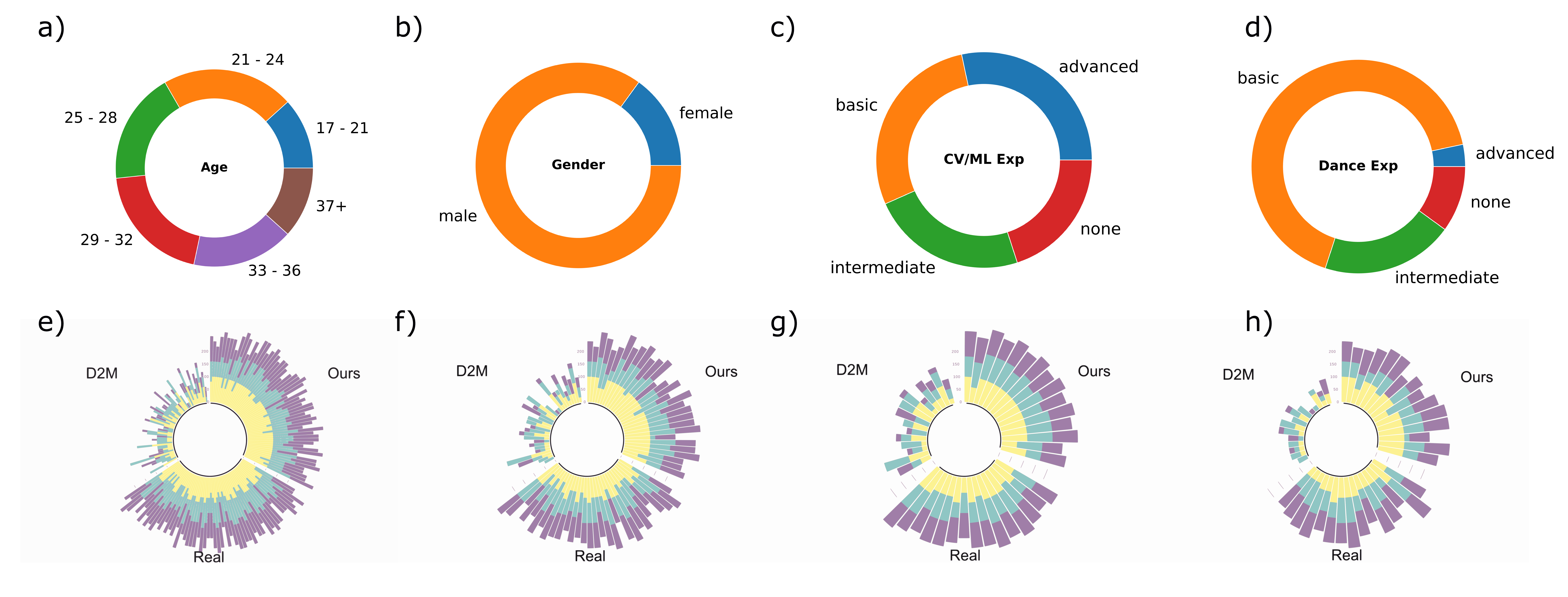}
    \caption{The plots a), b), c) and d) show the profile distribution of the participants of our user study; The plots e), f), g) and h) show the results of the study. In the plots of semi-circles are shown the results of the user evaluation; each stacked bar represents one user evaluation and the colors of each stacked bar indicates the dance styles (Ballet $=$ yellow, Michael Jackson (MJ) $=$ blue, and Salsa $=$ purple). \textbf{e)} We show the results for all $\mathbf{60}$ users that fully answered our study; \textbf{f)} Results for the users which achieved top $\mathbf{27\%}$ scores and the $\mathbf{27\%}$ which achieved the bottom scores; \textbf{g)} Results for the $\mathbf{27\%}$ user which achieved top scores; \textbf{h)} Results for the $\mathbf{27\%}$ users which achieved worst scores. 
    }
    \label{fig:user-study-profile}
\end{figure*}

We conducted a perceptual study with $60$ users and collected the age, gender, Computer Vision/Machine Learning experience, and familiarity with different dance styles for each user. Figure~\ref{fig:user-study-profile} shows the profiles of the participants.

The perceptual study was composed of $45$ randomly sorted tests. For each test, the user watches a video (with no sound) synthesized by vid2vid using a generated set of poses. Then we asked them to associate the motion performed on the synthesized video as belonging to one of the audio classes: Ballet, Michael Jackson, or Salsa. In each question, the users were supposed to listen to one audio of each class to help them to classify the video. 
The set of questions was composed of $15$ videos of movements generated by our approach, $15$ videos generated by D2M~\cite{lee2019dancing2music}, and $15$ videos of real movements extracted from our training dataset. We applied the same transformations to all data and every video had an avatar performing the motion with a skeleton with approximately the same dimensions. We split equally the $15$ videos shown between the three dance styles.

\begin{figure}[t!]
    \centering
    \includegraphics[width=0.95\linewidth]{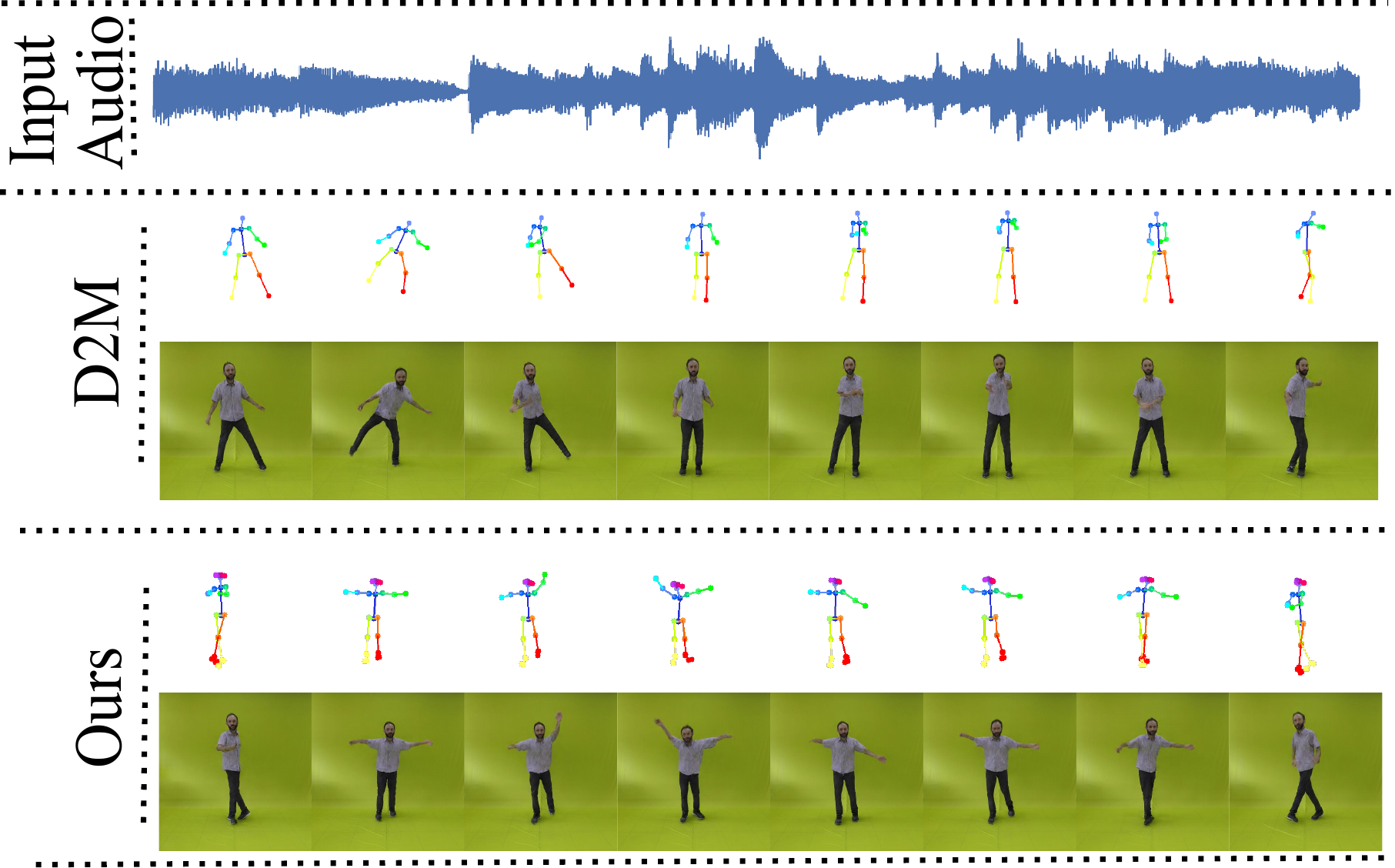}
    \caption{Results of our approach in comparison to D2M \cite{lee2019dancing2music} for \textit{Ballet}, the dance style shared by both methods.}
    \label{fig:ours-vs-d2m}
\end{figure}

\begin{figure*}[!h]
    \centering
    \includegraphics[width=0.72\linewidth]{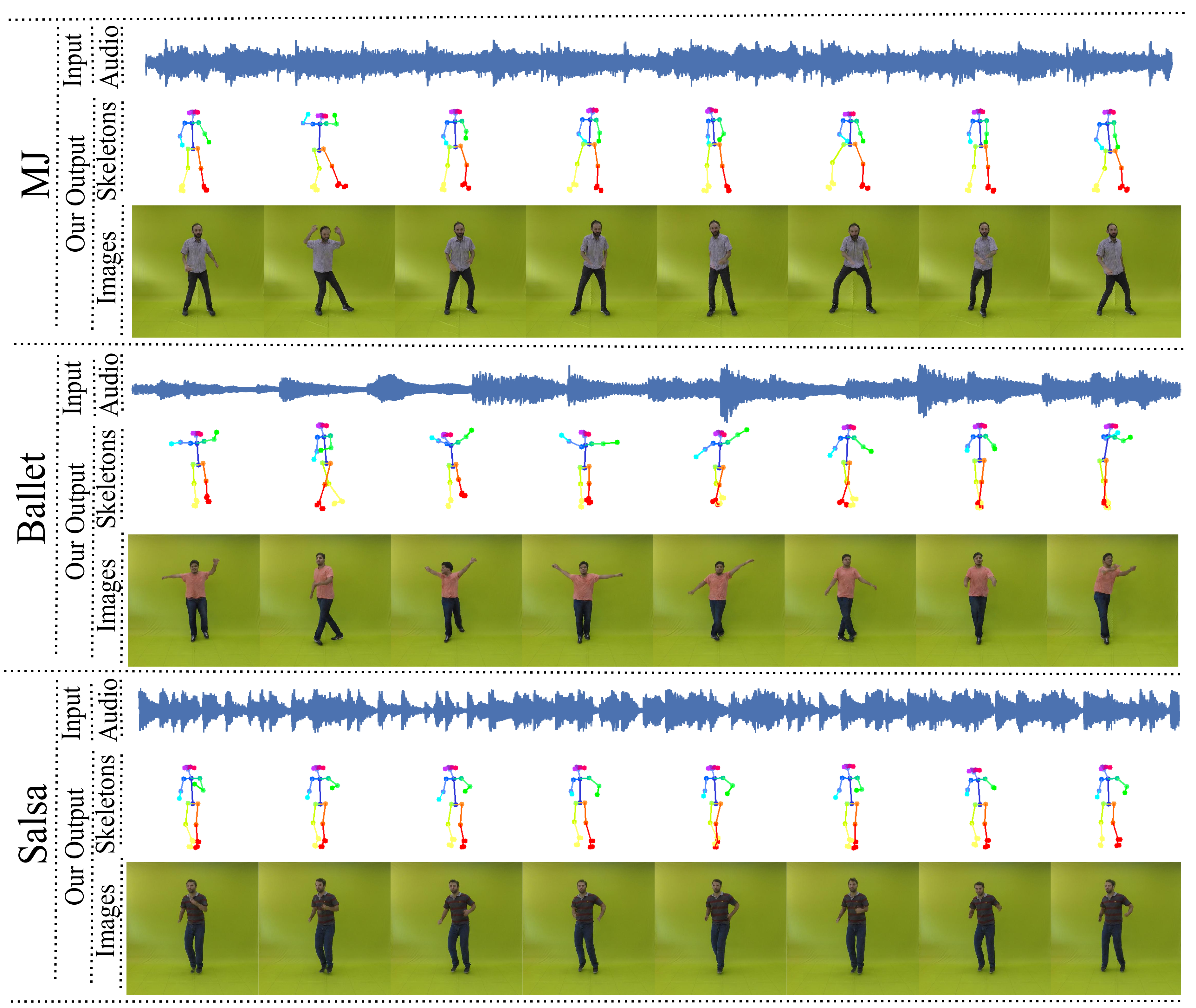}
    \caption{Qualitative results using audio sequences with different styles. In each sequence: First row: input audio; Second row: the sequence of skeletons generated with our method; Third row: the animation of an avatar by vid2vid using our skeletons.}
    \label{fig:ours-results}
    \vspace*{\floatsep}
    \centering
    \includegraphics[width=0.73\linewidth]{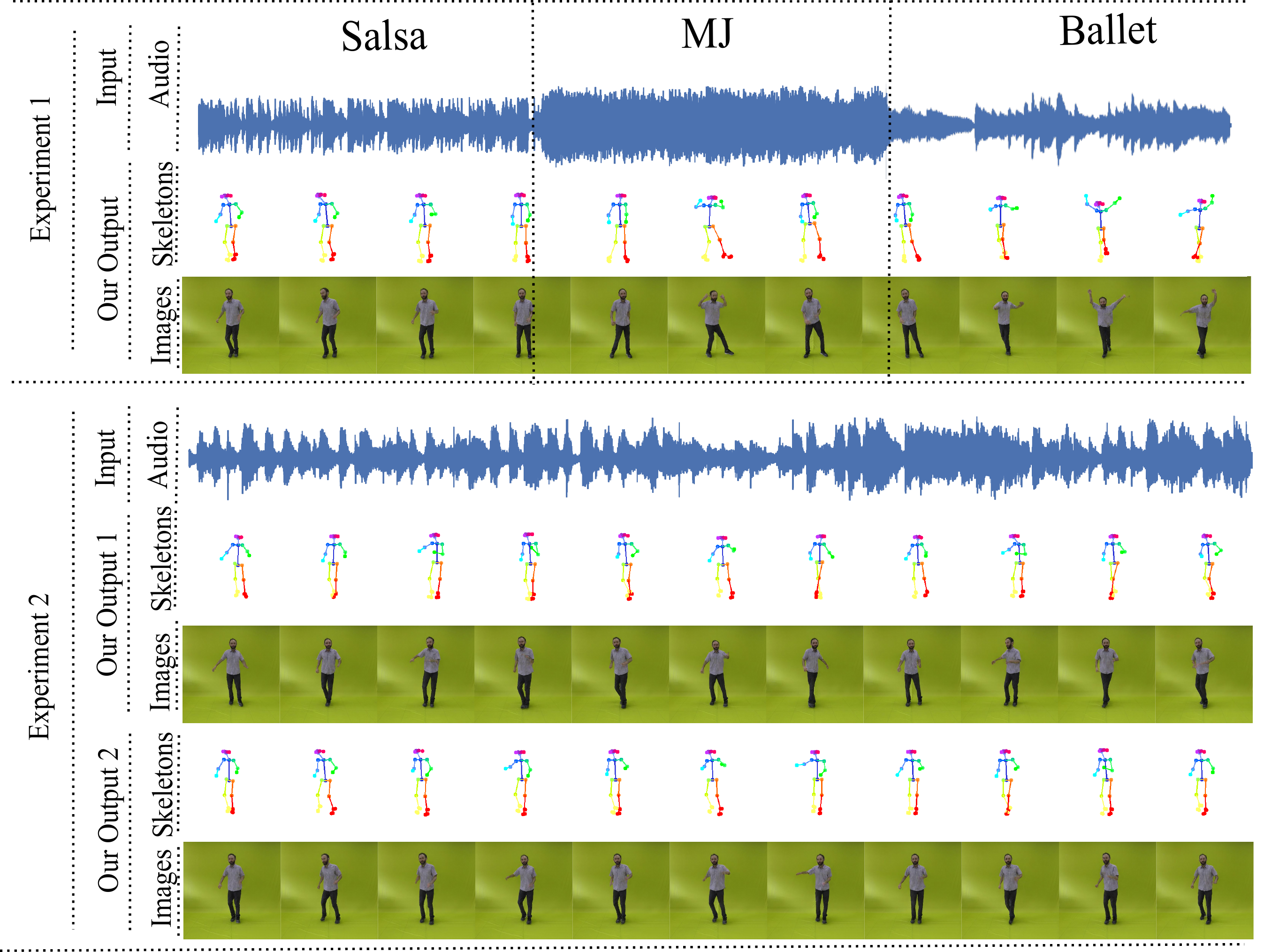}
    \caption{Experiment 1 shows the ability of our method to generate different sequences with smooth transition from one given input audio composed of different music styles. Experiment 2 illustrates that our method can generate different sequences from a given input music.}
    \label{fig:ours-results_2}
\end{figure*}

From Table~\ref{table:results_perceptual} and Figure~\ref{fig:user-study-profile}, we draw the following observations: first, our method achieved similar motion perceptual performance to the one obtained from real data. Second, our method outperformed the D2M method with a large margin. Thus, we argue that our method was capable of generating realistic samples of movement taking into account two of the following aspects: \textit{i)} Our performance is similar to the results from real motion data in a blind study; \textit{ii)} Users show higher accuracy in categorizing our generated motion. Furthermore, as far as the quality of a movement being individual is concerned, Figures~\ref{fig:ours-vs-d2m} and~\ref{fig:ours-results} show that our method was also able to generate samples with motion variability among samples.

We ran two statistical tests,  \textit{Difficulty Index} and \textit{Item Discrimination Index}, to test the validity of the questions in our study. The Difficulty Index measures how easy to answer an item is by determining the proportion of users who answered the question correctly, \ie, the accuracy. On the other hand, the Item Discrimination Index measures how a given test question can differentiate between users that mastered the motion style classification from those who have not. Our methodology analysis was based on the guidelines described by Luger and Bowles~\cite{LugerB16}. Table~\ref{table:results_perceptual} shows the average values of the indexes for all questions in the study. One can clearly observe that our method's questions had a higher difficulty index value, which means it was easier for the participants to answer them correctly and, in some cases, even easier than the real motion data. Regarding the discrimination index, we point out that the questions cannot be considered good enough to separate the ability level of those who took the test, since items with discrimination indexes values between $0$ and $0.29$ are not considered good selectors~\cite{ebel1991essentials}. These results suggest that our method and the videos obtained from real sequences look natural for most users, while the videos generated by \cite{lee2019dancing2music} were confusing.



\begin{table}[t!]
	\centering
	\caption{Quantitative metrics from the user perceptual study.}
	\label{table:results_perceptual}
	\resizebox{\linewidth}{!}{
		\begin{tabular}{@{}crrrrrrrrc@{}}
			\toprule 
			 & \multicolumn{3}{c}{Difficulty Index$^1$ } & \phantom{abc} & \multicolumn{3}{c}{Discrimination Index $^2$} \\ 
			 \cmidrule{2-4} \cmidrule{6-8} 
			\multicolumn{1}{r}{\textbf{Dance Style}}                      & D2M       & \multicolumn{1}{r}{Ours}    & \multicolumn{1}{r}{Real}&& D2M    & \multicolumn{1}{r}{Ours}       & \multicolumn{1}{r}{Real}\\ \midrule 
			\multicolumn{1}{l}{Ballet}    & $0.183$       & 0.943    &  \multicolumn{1}{r}{$\mathbf{0.987}$} && \multicolumn{1}{r}{$\mathbf{0.080}$}     & \multicolumn{1}{r}{$\mathbf{0.080}$}     &0.033\\
			\multicolumn{1}{l}{MJ}    & $0.403$       & 0.760    & \multicolumn{1}{r}{$\mathbf{0.843}$} & & $0.140$     & \multicolumn{1}{r}{$\mathbf{0.240}$}     &0.120 \\
			\multicolumn{1}{l}{Salsa}    & $0.286$       & \multicolumn{1}{r}{$\mathbf{0.940}$}    & 0.820 && $0.100$     & \multicolumn{1}{r}{0.030}    &\multicolumn{1}{r}{$\mathbf{0.180}$}  \\
			\multicolumn{1}{l}{\textit{Average}}    & $0.290$       & 0.881    & \multicolumn{1}{r}{$\mathbf{0.883}$} && $0.106$     & \multicolumn{1}{r}{$\mathbf{0.116}$}    &0.111  \\
			\multicolumn{1}{l}{}    & \multicolumn{3}{c}{\scriptsize{$^1$\textit{Better closer to 1.}}} & & \multicolumn{3}{c}{\scriptsize{$^2$\textit{Better closer to 1.}}} \\
			\bottomrule 
		\end{tabular}
	}
\end{table}


\subsection{Quantitative Evaluation}
\begin{table*}[t]
	\centering
	\caption{Quantitative evaluation according to FID, GAN-Train, and GAN-Test metrics.}
	\label{table:results_quantitative}
	\resizebox{\linewidth}{!}{
		\begin{tabular}{@{}crrrrrrrrrrrrc@{}}
			\toprule 
			& \multicolumn{3}{c}{FID$^1$ } & \phantom{abc} & \multicolumn{3}{c}{GAN-Train $^2$} & \phantom{abc} & \multicolumn{3}{c}{GAN-Test\* $^3$}\\ 
			\cmidrule{2-4} \cmidrule{6-8} \cmidrule{10-12}
			\multicolumn{1}{l}{\textbf{Dance Style}}                      & \multicolumn{1}{c}{D2M}       & \multicolumn{1}{c}{Ours}    & \multicolumn{1}{c}{Real}&& \multicolumn{1}{c}{D2M}    & \multicolumn{1}{c}{Ours}       & \multicolumn{1}{c}{Real}&& \multicolumn{1}{c}{D2M}    & \multicolumn{1}{c}{Ours}       & \multicolumn{1}{c}{Real}\\ \midrule 
			
			\multicolumn{1}{l}{Ballet}    & $20.20\pm4.41$       & \multicolumn{1}{r}{$\mathbf{3.18\pm1.43}$}    &$2.09\pm0.58$&& $0.36\pm0.15$     & \multicolumn{1}{r}{$\mathbf{0.89\pm0.10}$}     &$0.80\pm0.12$ && $0.07\pm0.04$     & \multicolumn{1}{r}{$\mathbf{0.80\pm0.14}$}     &$0.77\pm0.11$\\  
			\multicolumn{1}{l}{MJ}    & $\mathbf{4.38\pm1.94}$       & \multicolumn{1}{r}{$8.03\pm3.55$}    &$5.60\pm1.42$& & $0.34\pm0.15$     & \multicolumn{1}{r}{$\mathbf{0.60\pm0.13}$}     &$0.59\pm0.04$ && $\mathbf{0.70\pm0.14}$     & \multicolumn{1}{r}{$0.46\pm0.18$}     &$0.60\pm0.09$ \\
			\multicolumn{1}{l}{Salsa}    & $12.23\pm3.20$       & \multicolumn{1}{r}{$\mathbf{4.29\pm2.38}$}    & $2.40\pm0.75$ && $\mathbf{0.32\pm0.17}$     & \multicolumn{1}{r}{$0.31\pm0.11$}    &$0.50\pm0.16$   && $0.26\pm0.14$       & \multicolumn{1}{r}{$\mathbf{0.96\pm0.11}$}    & $0.90\pm0.06$   \\
			\multicolumn{1}{l}{\textit{Average}}    & $12.27\pm7.27$       & \multicolumn{1}{r}{$\mathbf{5.17\pm3.33}$}    & $3.36\pm1.86$ && $0.34\pm0.16$     & \multicolumn{1}{r}{$\mathbf{0.60\pm0.26}$}    &$0.63\pm0.17$  && $0.34\pm0.29$       & \multicolumn{1}{r}{$\mathbf{0.74\pm0.25}$}    & $0.76\pm0.15$  \\

			\multicolumn{1}{l}{}    & \multicolumn{3}{c}{\scriptsize{$^1$\textit{Better closer to 0.}}} & & \multicolumn{3}{c}{\scriptsize{$^2$\textit{Better closer to 1.}}} & & \multicolumn{3}{c}{\scriptsize{$^3$\textit{Better closer to 1.}}}\\
			\bottomrule 
		\end{tabular}
	}
\end{table*}

For a more detailed performance assessment regarding the similarity between the learned distributions and the real ones, we use the commonly used Fr\'echet Inception Distance (FID). We computed the FID values using motion features extracted from the action recognition ST-GCN model presented in~\cite{yan2018spatial}, similar to the metric used in \cite{yan2019convolutional,lee2019dancing2music}.
We train the ST-GCN model $50$ times using the same set of hyperparameters. The trained models achieved accuracy scores higher than $90\%$ for almost all $50$ training trials. The data used to train the feature vector extractor was not used to train any of the methods evaluated in this paper. Table~\ref{table:results_quantitative} shows the results for the FID metric.



We also computed the GAN-Train and GAN-Test metrics, two well-known GAN evaluation metrics~\cite{Shmelkov18}. To compute the values of the GAN-Train metric, we trained the ST-GCN model in a set composed of dance motion samples generated by our method and another set with generated motions by D2M. Then, we tested the model in the evaluation set (real samples). The GAN-Test values were obtained by training the same classifier in the evaluation set and tested in the sets of generated motions. For each metric, we ran $50$ training rounds and reported the average accuracy with the standard deviation in Table~\ref{table:results_quantitative}. Our method achieved superior performance as compared to D2M. 

We can also note that the generator performs better in some dance styles. Since some motions are more complicated than others, the performance of our generator can be better synthesizing less complicated motions related to a particular audio class related to a dance style. For instance, the Michael Jackson style contains a richer set of motions with the skeleton joints rotating and translating in a variety of configurations. The Ballet style, on the other hand, is composed of fewer poses and consequently, easier to synthesize.




\subsection{Qualitative Evaluation}

Figures~\ref{fig:ours-vs-d2m}, \ref{fig:ours-results}, and \ref{fig:ours-results_2} show some qualitative results. We can notice that the sequences generated by D2M presented some characteristics clearly inherent to the dance style, but they are not present along the whole sequence. For instance, in Figure~\ref{fig:ours-vs-d2m}, one can see that the last generated skeleton/frame looks like a spin, usually seen in ballet performances, but the previous poses do not indicate any correlation to this dance style. Conversely, our method generates poses commonly associated with ballet movements such as rotating the torso with stretched arms.

Figure~\ref{fig:ours-results} shows that for all three dance styles, the movement signature was preserved. Moreover, the \textit{Experiment 1} in Figure~\ref{fig:ours-results_2} demonstrates that our method is highly responsive to audio style changes since our classifier acts sequentially on subsequent music portions. This enables it to generate videos where the performer executes movements from different styles. Together these results show that our method holds the ability to create highly discriminative and plausible dance movements. Notice that qualitatively we outperformed D2M for all dance styles, including for the Ballet style, which D2M was carefully coined to address. \textit{Experiment 2} in Figure~\ref{fig:ours-results_2} also shows that our method can generate different sequences from a given input music. Since our model is conditioned on the music style from the audio classification pipeline, and not on the music itself, our method exhibits the capacity of generating varied motions while still preserving the learned motion signature of each dance style.

%% file: conclussion.tex
\section{Conclusion}

In this paper, we propose a new method for synthesizing human motion from music. Unlike previous methods, we use graph convolutional networks trained using an adversarial regime to address the problem. We use audio data to condition the motion generation and produce realistic human movements with respect to a dance style. We achieved qualitative and quantitative performance as compared to state of the art. Our method outperformed Dancing to Music in terms of FID, GAN-Train, and GAN-Test metrics. We also conducted a user study, which showed that our method received similar scores to real dance movements, which was not observed in the competitors.

Moreover, we presented a new dataset with audio and visual data, carefully collected to train and evaluate algorithms designed to synthesize human motion in dance scenario. Our method and the dataset are one step towards fostering new   approaches for generating human motions.

As future work, we intend to extend our method to infer 3D human motions, which will allow us to use the generated movements in different animation frameworks. We also plan to increase the dataset size by adding more dance styles.

\paragraph*{\bf Acknowledgements}

The authors thank CAPES, CNPq, and FAPEMIG for funding this work. We also thank NVIDIA for the donation of a Titan XP GPU used in this research.